\documentclass[authoryear,preprint,12pt]{elsarticle}

\usepackage{graphicx}
\usepackage{amssymb}
\usepackage{wasysym}
\usepackage{lineno}
\usepackage{rotating}
\usepackage{changepage}
\usepackage[usenames]{color}
\RequirePackage[colorlinks]{hyperref}

\RequirePackage{hypernat}

\journal{Earth and Planetary Science Letters}

\begin{document}

\begin{frontmatter}



\title{A phase-space model for Pleistocene ice volume}

\author[uva,ias]{John Z. Imbrie\corref{cor1}}
\ead{imbrie@virginia.edu}
\author[sas]{Annabel Imbrie-Moore}
\ead{aimbriemoore@gmail.com}
\author[ucsb]{Lorraine E. Lisiecki\fnref{fn1}}
\ead{lisiecki@geol.ucsb.edu}
\cortext[cor1]{Corresponding author}
\fntext[fn1]{Support provided by NSF-MGG 0926735}
\address[uva]{Department of Mathematics,
University of Virginia,
Charlottesville, VA 22904, USA}
\address[ias]{School of Mathematics,
Institute for Advanced Study,
Princeton, NJ 08540, USA}
\address[sas]{St. Andrew's School, 350 Noxontown Rd., Middletown, DE 19709, USA}
\address[ucsb]{Department of Earth Science, University of California, Santa Barbara, CA 93106, USA}
\begin{abstract}
Abstract: We present a phase-space model that simulates Pleistocene
ice volume changes based on Earth's orbital parameters. Terminations
in the model are triggered by a combination of ice volume and orbital
forcing and agree well with age estimates for Late Pleistocene
terminations. The average phase at which model terminations begin is
approximately 90$\pm$90$^\circ$ before the maxima in all three orbital
cycles. The large variability in phase is likely caused by
interactions between the three cycles and ice volume. 
Unlike previous
ice volume models, this model 
produces an orbitally driven increase in 100-kyr power
during the mid-Pleistocene transition without any change in model
parameters. This supports the hypothesis that Pleistocene variations
in the 100-kyr power of glacial cycles could be caused,
at least in part,
by changes in
Earth's orbital parameters, such as amplitude modulation of the
100-kyr eccentricity cycle, rather than changes within the climate
system.

\end{abstract}


\end{frontmatter}

 %\linenumbers

\section{Introduction}
\label{introduction}
Numerous studies have demonstrated that Pleistocene glacial cycles are
linked to cyclic changes in Earth's orbital parameters \citep{Hay76,Imb92,Lis07};
%[Hays et al.,  1976; Imbrie et al., 1992; Lisiecki and Raymo, 2007]; 
however, many
questions remain about how orbital cycles in insolation produce the
observed climate response. The most contentious problem is why
late Pleistocene climate records are dominated by 100-kyr
cyclicity. Insolation changes are dominated by 41-kyr obliquity and
23-kyr precession cycles whereas the 100-kyr eccentricity cycle
produces negligible 100-kyr power in seasonal or mean annual
insolation. Thus, various studies have proposed that 100-kyr glacial
cycles are a response to the eccentricity-driven modulation of
precession \citep{Ray97,Lis10b}, bundling of obliquity cycles
\citep{Huy05,Liu08},
%[Huybers and Wunsch, 2005; Liu et al., 2008], 
and/or internal
oscillations \citep{Sal84,Gil00,Tog08}
% [Saltzman et al., 1984; Gildor and Tziperman, 2000;Toggweiler, 2008].

A closely related problem is the question of why the dominant glacial
cycle shifted from 41 kyr to 100 kyr during the mid-Pleistocene.  Most
commonly, this mid-Pleistocene transition (MPT) is attributed either
to a drop in atmospheric CO$_2$ levels \citep{Ray97,Pai98,Hon09}
% [Raymo, 1997; Paillard, 1998; Honisch et al., 2009] 
or erosion of the continental regolith \citep{Cla06}, but other mechanisms have also been proposed. Some
hypotheses do not require any internal changes in the climate system,
attributing the shift to chaotic or irregular mode-switching \citep{Sal93,Huy09}, or to
a change in the character of insolation forcing \citep{Lis10b}. For example, the 100-kyr power of the climate response is
observed to be negatively correlated with the 100-kyr power of
eccentricity for at least the last 3 Myr \citep{Lis10b,Mey10}.

Many simple models have produced 100-kyr cycles as responses to
precession and/or obliquity using different nonlinear responses to
orbital forcing. Early models had difficulty reproducing the large
amplitude of Marine Isotope Stage (MIS) 11 during weak orbital forcing
and produced too much 400-kyr power \citep[\textit{e.g.}][]{Imb80}.
However, newer multi-state models have solved these particular
problems \citep{Pai98,Par03}. Many models
can also reproduce a transition from 41-kyr to 100-kyr cyclicity
during the mid-Pleistocene transition by changing certain model
parameters or climate boundary conditions \citep{Ray97,Pai98,Ash04,Pai04,Cla06,Huy07} 
%[Raymo, 1997; Paillard, 1998; Ashkenazy and Tziperman, 2004; Clark et al., 2006; Huybers, 2007].
In fact, the wide variety of ways in which 100-kyr glacial
cycles can be produced makes it difficult to determine which, if any,
of the models correctly describes the source of 100-kyr glacial
cyclicity \citep{Tzi06}.

We present a new, phase-space model of Pleistocene ice volume that
generates 100-kyr cycles in the Late Pleistocene as a response to
obliquity and precession forcing. Like \citet{Par03}, we use a threshold for glacial terminations. However, ours is a phase-space threshold: a function of ice volume and its rate of change. 
Our model is 
the first to produce an orbitally driven
increase in 100-kyr power
during the mid-Pleistocene transition without any change in model
parameters. In section \ref{methods}, we describe the model and the derivation of
its parameters. In section \ref{results}, we compare the model results and climate
data for the last 3 Myr, with emphasis on the timing of 100-kyr
glacial terminations and changes in spectral power. In section \ref{discussion}, we
discuss (1) parameterization of the relationships between ice volume
and orbital forcing, (2) the timing of terminations with respect to
orbital forcing, and (3) the mid-Pleistocene transition. Finally,
section \ref{conclusions} summarizes our conclusions.

\section{Methods}
\label{methods}
\subsection{Overview}
\label{overview}
We use a statistical analysis of the ice-volume record to guide the development of a set of evolution equations which accurately model the dynamics of glacial cycles. This is an important shift in perspective, in that we are not testing specific physical mechanisms that may be responsible for key features of the record. Rather, we hope that the form of the equations will help clarify the discussion of possible mechanisms. Assuming that ice volume is the slowest mode in the climate system, we look for equations involving a single variable $y$ and an orbital forcing term, and then explain as much of the low-frequency variation in the record as possible.
\subsection{Variables}
\label{variables}
Development and parameterization of the model is guided by analysis of the LR04 global stack of benthic $\delta^{18}$O \citep{Lis05}, which is a proxy for global ice volume and deepwater temperature.
The LR04 stack from 0 to 1500 ka is taken with a sign change, as is conventional so that larger values correspond to warmer epochs (smaller ice volume). The stack is interpolated as needed to get a record sampled every thousand years. As we are interested in the slowly varying aspects of the record, we put the data through a Gaussian notch filter centered at 0 with a bandwidth of .1 
$\mathrm{kyr}^{-1}$. The result was standardized by subtracting the mean and dividing by the standard deviation, producing a function $y(t)$ for the ice-volume record over time scaled to run roughly from -2 to 2. 

Combinations of orbital functions $\varepsilon$ (obliquity), $e \sin \omega$ (precession) and $e \cos \omega$ (phase-shifted precession) are used as a forcing for our model, and are taken from \citep{Las04}. Insolation at most latitudes and seasons can be represented quite accurately by a combination of these three orbital functions. The variables were standardized based on the mean and standard deviation from 0 to 1500 ka, with a thousand year sampling interval. The resulting variables are denoted $obl, esi, eco$.

\subsection{Phase-space picture}
\label{phasespacepicture}
\begin{figure}
\noindent
\center
\includegraphics[width=\textwidth]{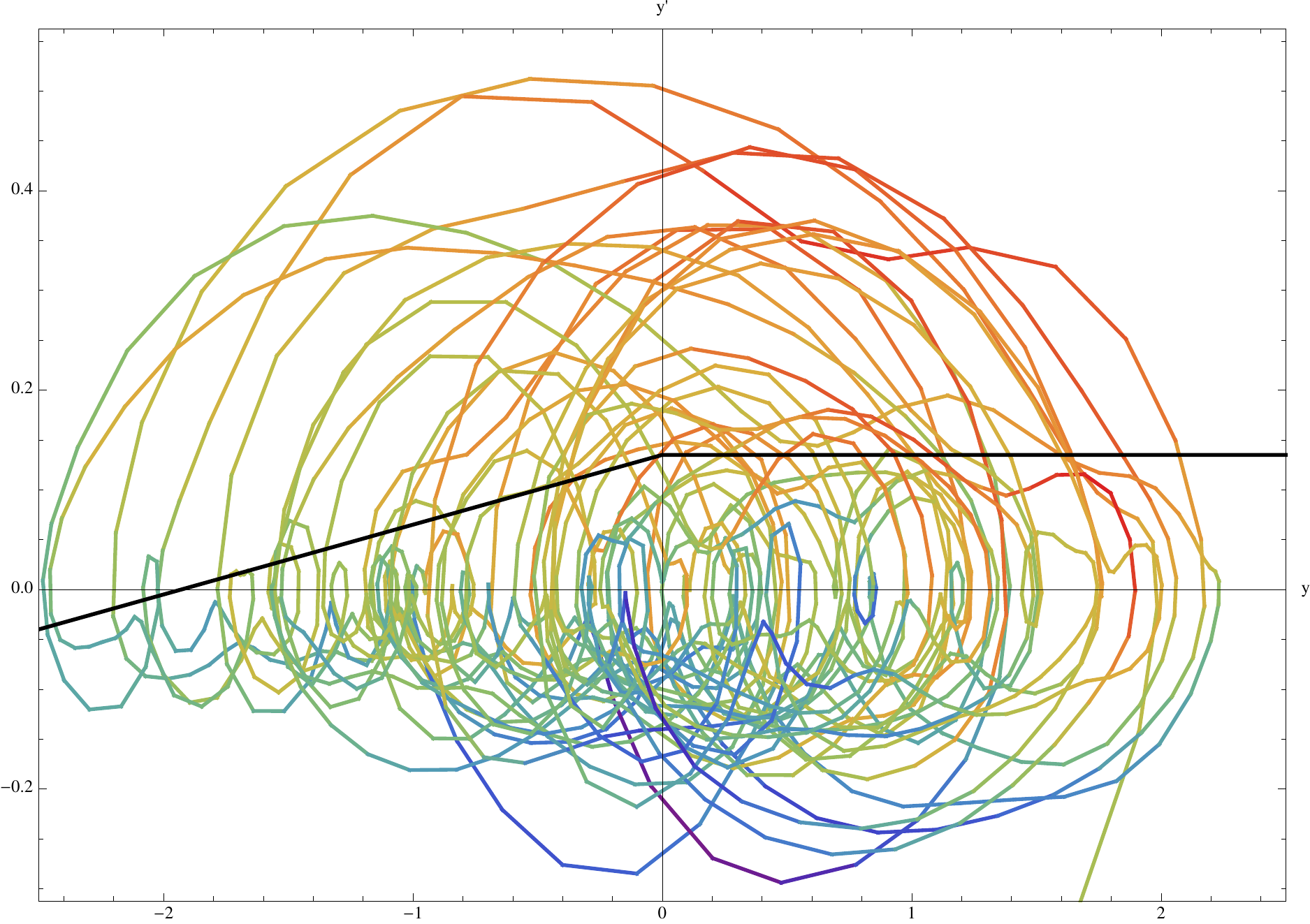}
\caption{Phase-space trace of the filtered ice-volume record. Colors indicate forcing. Black line indicates the function $\textrm{trans}(x)$ which delineates the model's boundary between the accumulation state and the termination state. 
(See supplementary online video.)
}
 \label{squiggle}
 \end{figure}

We examine the ice-volume record in a 2-dimensional plot, with $y$ on the horizontal axis and $y'$ (the time rate of change of $y$) on the vertical axis. Positive values denote $y' > 0$ (warming epochs). Colors denote the forcing function (discussed below), with red for large positive (warming) and blue for large negative (cooling). One thing that becomes clear from Figure \ref{squiggle} is the special role that the large semicircular loops play in the ice-volume dynamics. These are traversed in a clockwise fashion and correspond to periods of rapid warming (terminations). Terminations have long been recognized as key features of the ice-volume record. Most of the time, however, the system remains fairly close to the horizontal axis, moving gradually left and bobbing up and down as it adapts slowly to changes in forcing.

\subsection{Threshold for terminations}
\label{thresholdforterminations}
In Figure \ref{squiggle}  one can observe just above the horizontal axis that there is a transition zone where some loops head back to the axis and others head upwards to initiate a large loop. Evidently, if the rate of warming is large enough, the climate will shift into a termination mode with runaway melting. The threshold appears to be a diagonal line extending roughly from (-2,0) to (0,.13). This leads us to define a transition line 
\begin{equation}
\mathrm{trans}(x)=\min\{.135+.07x,.135\}
\label{trans}
\end{equation}
(plotted onto Figure \ref{squiggle}) which incorporates this diagonal line and extends it horizontally to the right of the vertical axis. Above the line, we consider the system to be in termination state; below it is in glaciating state. Multi-state models have been employed with considerable success by \citet{Par03} using a threshold which depends on a combination of insolation and ice volume. Here the diagonal line represents a combination of $y' $ and $y$ (ice volume)---but insofar as $y'$ and insolation are correlated, the concepts are similar. 

\subsection{Accumulation state}
\label{accumulationstate}
The accumulation state is modeled with a first-order differential equation
 \begin{equation}
y'=g(y)+F(y,t),
\label{yprime}
\end{equation}
where $g(y)$ describes the internal tendency of the ice sheets to grow or retreat, and $F(y,t)$ describes the external forcing.
A first-order equation is a natural starting point for a system responding to a variable heat source. Additionally, the phase lag with respect to obliquity is characteristic of a first-order equation. As we will see, there is significant variation in $y$ of the sensitivity of ice volume  to various components of orbital forcing. This is the reason for allowing the forcing function to depend on $y$ as well as $t$.

Clues for what might be reasonable choices for $g$ were obtained by a weighted regression analysis. For the period 1500-0 ka, a weighted least squares regression was performed for $y'$ with $obl$, $esi$, $eco$ used as explanatory variables. The following weights were used: $\exp[-(y(t)-y_0)^2/d^2]$,
with $d=.45$. In this way, we were able to get an understanding of the contributions of the individual forcing functions to $y'$. Furthermore, we could learn how the contributions vary with ice volume $y_0$ by effectively restricting attention to data points with $y$ near $y_0$. The linear model looks as follows:
\begin{equation}
y'=h_1(y_0)obl+h_2(y_0)esi+h_3(y_0)eco+r,
\label{yprimelinear}
\end{equation}
where $h_1, h_2, h_3$ are the regression coefficients and $r$ is the residual.
The residual $r(t)$ has considerable scatter but by averaging the values of $r$ using a weight function as above, we obtained an estimate for the internal evolution function:
\begin{equation}
g(y_0)=\frac{1}{N} \sum_t r(t)\exp[-(y(t)-y_0)^2/d^2],
\label{gequation}
\end{equation}
where
\begin{equation}
N=\sum_t \exp[-(y(t)-y_0)^2/d^2],
\label{N}
\end{equation}

\begin{figure}
\noindent
\begin{adjustwidth}{-2.5cm}{-2.5cm}
\center
\includegraphics[width=36pc]{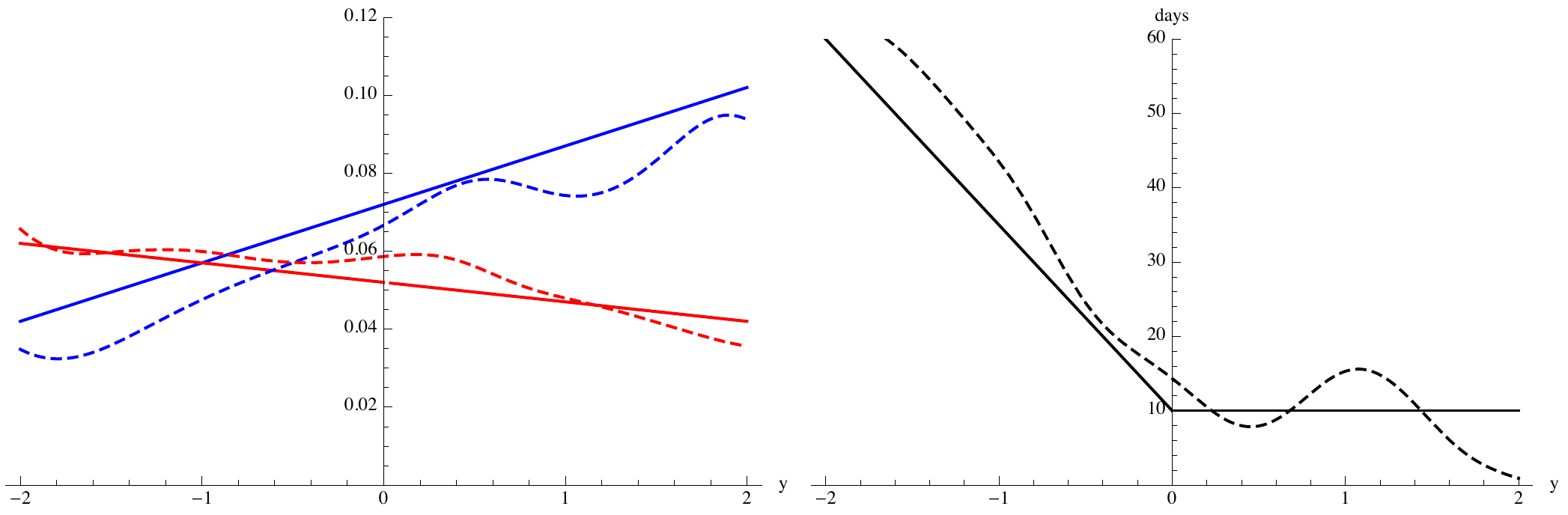}
\end{adjustwidth}
\caption{Parameters in the forcing function. On the left, dashed lines indicate the original coefficients $h_1(y)$ (blue---obliquity) and $A(y)$ (red---precession amplitude) obtained from the data by a weighted regression. Solid lines indicate the straight-line versions used in the model. On the right, the dashed line indicates the precession phase angle, as determined by regression. Solid line indicates the piecewise-linear version used in the model.}
 \label{2a2b}
 \end{figure}

\begin{figure}
\noindent
\center
\includegraphics[width=22pc]{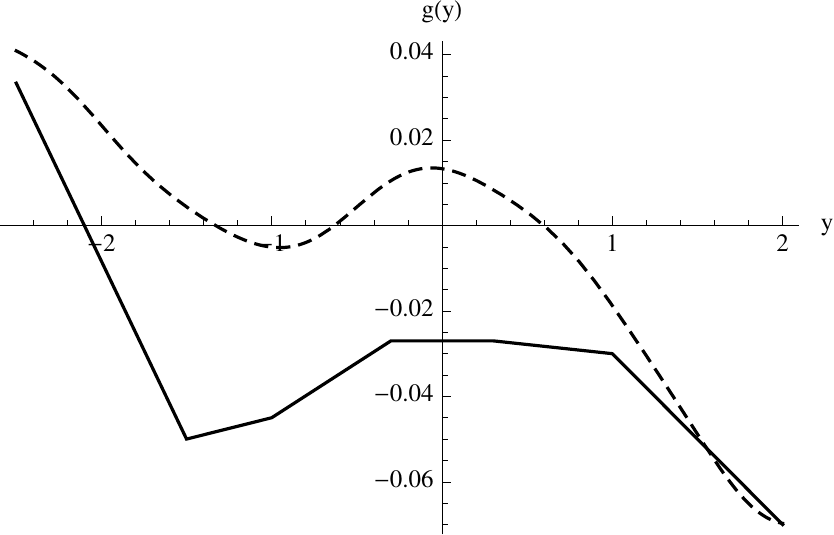}
\caption{Dashed line indicates the drift function $g(y)$ obtained from smoothed residuals. Solid line indicates the piecewise-linear version used in the model.}
\label{g}
\end{figure}

In formulating the model, we used these functions $h_1, h_2, h_3, g$ as guidelines only, giving ourselves the freedom to tweak the curves to improve the results. In our model, the coefficients $h_1$ and  $A=\sqrt{h_2^2+h_3^2}$ were obtained from straight-line versions of the corresponding regression coefficients. See Figure \ref{2a2b} (left).
We used a piecewise linear approximation for the phase angle:
 \begin{equation}
days=\frac{180}{\pi} \arctan\left(\frac{h_3}{h_2}\right) \approx \max\{10,10-25y\}.
\label{days}
\end{equation}
This corresponds to the number of days prior to June 21 for the reference point of the precession angle (or for the date in an equivalent insolation curve). See Figure \ref{2a2b} (right).
Model $h_1$, $A$, and $days$ curves readily translate into a forcing function
\begin{equation}
F(y,t)=h_1(y(t))obl(t)+h_2(y(t))esi(t)+h_3(y(t))eco(t).
\label{F}
\end{equation}
Lastly, we compare in Figure \ref{g} the original $g(y)$ with the piecewise-linear version used in our model. 
Note that there is a similarity in shape, but the version used for the accumulation state is lower. This is due to the fact that the original $g(y)$ came from averaging all the data, including the terminations, which have large positive values of $y'$. When one restricts attention to data in the accumulation state, a lower curve should be expected. 

One can visualize the effect of the forcing $F(y,t)$ as moving this curve up and down. Although this picture is typical, for some (positive) values of the forcing, there are two stable points, and for still larger values there is just one stable point on the right (deglaciated). 
\subsection{Terminations}
\label{terminations}
Inspection of the phase-space trace in Figure \ref{squiggle} leads to the conclusion that a first-order equation is inadequate to describe the system above the black line. In particular, there is evidence of inertia, \textit{i.e.} a tendency for $y'$ to be consistent over successive time intervals. Therefore, an accurate description of the time evolution requires a second-order differential equation, which can be visualized as a flow in phase space.

Figure \ref{squiggle} shows
 that once above the line $\textrm{trans}(x)$ the system moves in a roughly semicircular pattern until it returns to the vicinity of the horizontal axis. Approximately 12,000 years are required to traverse the semicircle. The endpoint of the semicircle is variable, but appears to correlate with the forcing
$F(y,t)$.
We define a semicircle destination function $d(y,t)$ by scaling $F$ linearly:
\begin{equation}
d(y,t)=-1+3(F(y,t)+.28)/.51.
\label{d}
\end{equation}
The result tends to lie between -1 and 2. The warmer the forcing, the further to the right the semicircle aims.

\begin{figure}
\noindent
\center
\includegraphics[width=25pc]{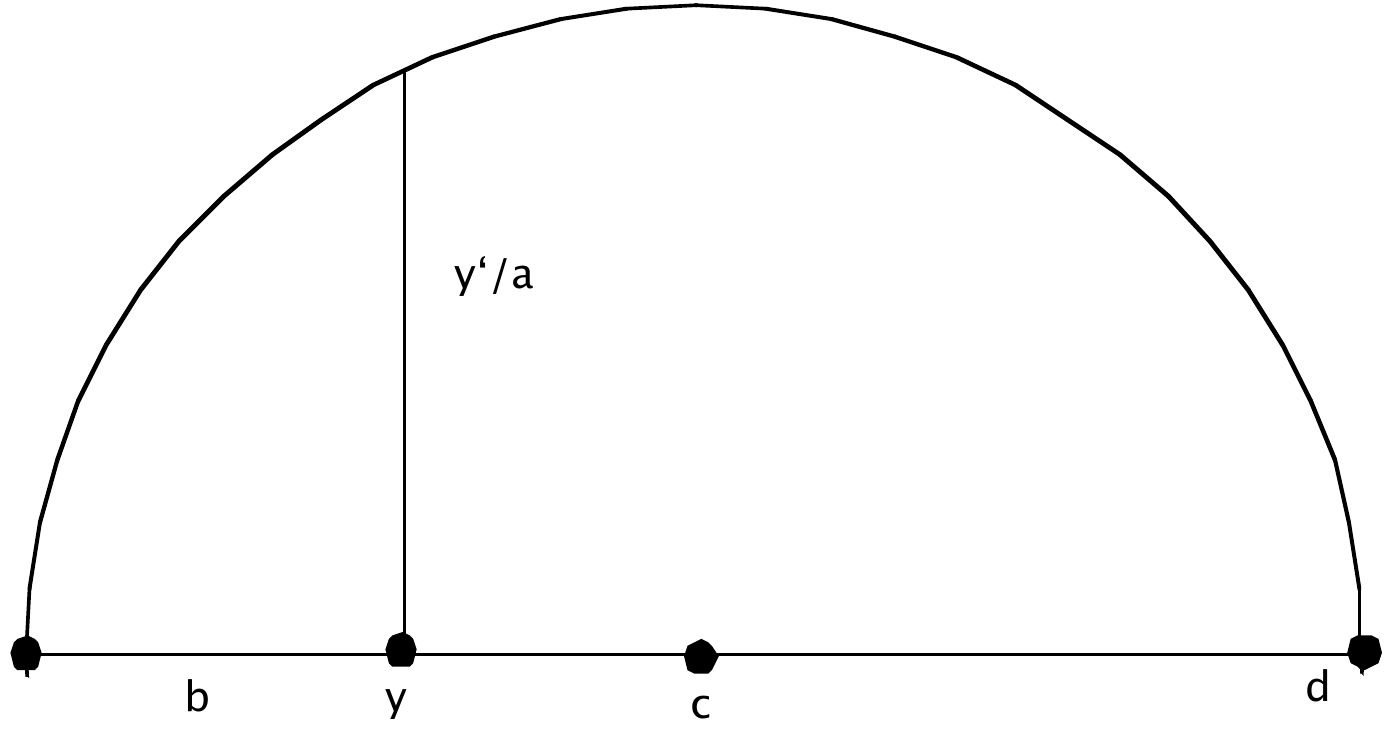}
\caption{Deriving the differential equation for terminations.}
\label{semicircle}
 \end{figure}

We find a differential equation which describes in phase space a semicircular flow with time-dependent destination. In Figure \ref{squiggle}, the vertical scale was expanded by a factor of 4 relative to the horizontal scale. Therefore, if we let the vertical coordinate be $y'/a$ with $a = 1/4$, we should see semicircular flow. The equation
$y''=-a^2(y-c)$
describes circular flow with center $c$ and period $2\pi/a$, which coincides with the desired 12,000 year semicircle traverse time. In order to express $c$ in terms of $d$, $y$, and $y'$, use some elementary geometry with Figure \ref{semicircle}.
Noting that $b(d-y)=(y'/a)^2$  and $c=(d+y-b)/2$,  we see that the equation can be written as
\begin{equation}
y''=\frac{1}{2}\left[a^2(d-y)-\frac{y'^2}{d-y}\right].
\label{ydoubleprime}
\end{equation}
A discretized version of this equation is used to generate the flow of our model while it is in the termination state (keep in mind that $d=d(y,t)$ from equation \ref{d}).

The system returns to the accumulation state as soon as $y'<\mathrm{trans}(y)$. In view of the time-dependence of $d$, it can happen that $y$ overshoots $d$, that is, $y > d$. In this case, $y'$ is set to 0. The discretized equation can run into problems when $y$ is close to $d$ due to the small denominator. This is handled by replacing any negative $y'$ with 0.

Physically, the underlying mechanism must be something which causes a continuation or even acceleration of melting once a threshold rate is passed. Under the right conditions isostatic depression can accentuate melting, and some models predict a catastrophic disintegration of the ice sheet \citep[\textit{e.g.}][]{Pel87}. Additional instability may arise if components of the ice sheet are marine-based; see the discussion in \citet{Den10}.

\subsection{Behavior of the model}
\label{behaviorofthemodel}
Due to the nature of threshold models, it is important to note that the model can be very sensitive to changes in parameters or in its position in phase space. The model can be close to the threshold for a termination and pushed over the edge, or a termination can be lost if the model is pulled back from the threshold. This is likely a feature of the real climate system as well. There is also an increase in sensitivity due to the way the forcing is dependent on the ice volume $y$. For example, ice-volume loss can be amplified if it causes an increase in the coefficient of a forcing function which happens to be positive or a decrease in the coefficient of a forcing function which happens to be negative.

\begin{sidewaysfigure}
\noindent
\center
\includegraphics[width=\textheight]{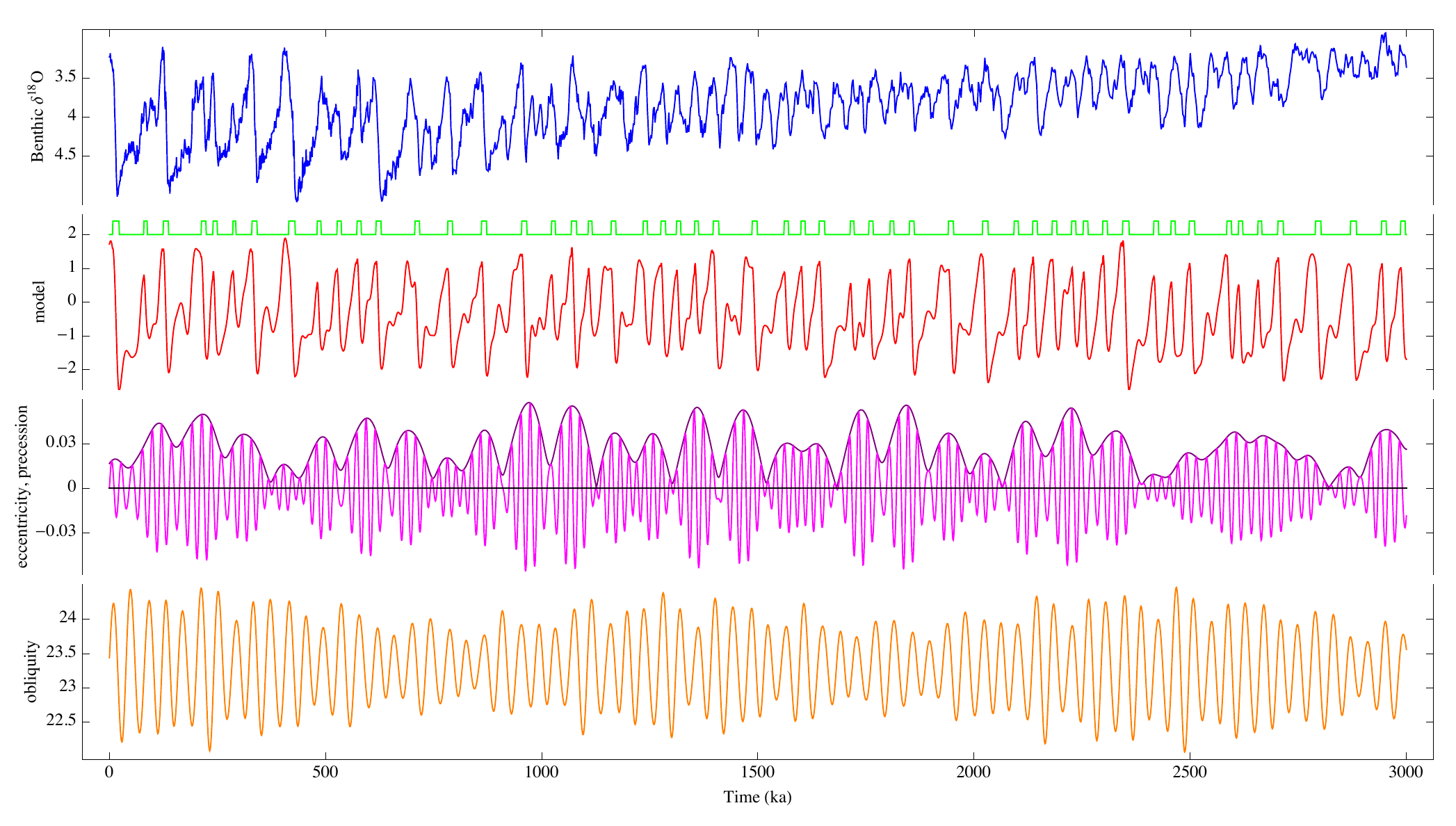}
\caption{Benthic $\delta^{18}$O (blue---not detrended) and phase-space model (red) are shown over the last 3000 ka. Model terminations are shown in green. Model precession forcing, its eccenctricity envelope, and obliquity are also shown.}
\label{model3000}
\end{sidewaysfigure}

One can get a general feel for the behavior of the model by observing that there are 40-kyr epochs where there is a termination with each obliquity cycle, as well as 100-kyr epochs where the model skips some obliquity cycles when it fails to reach the threshold for terminations. The idea of skipped obliquity cycles appears in \citet{Huy05}. Skipped cycles tend to happen during times of low eccentricity, since the combination of large precession and obliquity forcing will almost always push the model over the threshold \citep{Lis10b}.
Indeed, the largest terminations in our model occur soon after 400-kyr eccentricity minima---see Figure \ref{model3000} at 430 ka and 2360 ka. One of the important successes of our model is its accurate picture of stage 11 and termination V. (The model of \citet{Par03} also did well with this part of the record.)

The model also exhibits an increased sensitivity to the external forcing when ice volume is large, which means the system is primed to terminate after 80 to 100 kyr of accumulation (the exact timing of the termination is determined by the forcing). This leads to synchronization or phase-locking with eccentricity cycles when the model is in the 100-kyr mode. See \citet{Tzi06} for a discussion of this concept.

A notable success of our model is its ability to exhibit the mid-Pleistocene transition from 40-kyr to 100-kyr cycles without time-dependent parameters. The underlying mechanism is the decrease in the average eccentricity level about 900,000 years ago. One can see a confirmation that average eccentricity level is driving the switches between 40-kyr and 100-kyr mode by looking at the period of low eccentricity around 2500 ka which results in an increase in 100-kyr power in our model output, in agreement with the ice-volume record. See Figure \ref{model3000}, which shows the behavior of the model over the last 3000 kyr.

While second-order differential equations are useful in generating oscillatory solutions, they normally have a phase lag between 90$^\circ$ and 180$^\circ$, depending on the amount of damping
(imagine swinging a ball on a string).
The actual phase lag between typical insolation curves and the oxygen isotope record is close to, or a bit less than 90$^\circ$, which is typical of a first-order equation. While our model is second-order and has a propensity to oscillate when driven, it is first-order during the accumulation state, which is most of the time. 
Furthermore, the timing of terminations is determined in the accumulation state.
This enables us to maintain a good phase relationship between forcing and model output.

Our model is not equipped to handle the decrease in amplitude that becomes apparent when the data is viewed back to 3000 ka. Nevertheless, there is a fair amount of similarity in features. For both the model and the data, the 100-kyr cycles are most pronounced soon after the eccentricity minima at 2900, 2400, and 2100 ka (although the correspondence is imperfect).

\section{Results}
\label{results}
\subsection{Model-data comparison for 1.5--0 Ma}
\label{modeldatacomparison}
Model results are compared to a detrended version of the LR04 $\delta^{18}$O
stack from 3-0 Ma (Figure \ref{model3000}). Long-term trends in the mean and
variance of the stack are removed because the model includes no
provisions to simulate these changes. The stack is detrended first by
subtracting a best-fit fifth-order polynomial for 5.3-0 Ma and then
multiplying by $e^{0.34t}$, where t is time in Ma, to remove the
stack's exponential trend in variance \citep{Lis07}. Note
the stack does not provide definitive information about the ages of
ice volume change because its age model is orbitally tuned. However,
stack's tuned age model agrees within error with an untuned,
constant-sedimentation rate age model \citep{Lis10b} and with
radiometric age estimates of sea level change \citep{Tho06} and
magnetic reversals \citep[\textit{e.g.}][]{Can95}. Additionally, the
tuned age model assumes some variability in the phase of climate
response: (1) a linear increase in response time from 3--1.5 Ma and
(2) short-term phase changes at times when strict tuning would require
large deviations in global mean sedimentation rates \citep{Lis05}.

\begin{figure} 
\noindent
\begin{adjustwidth}{-3cm}{-3cm}
\center
\includegraphics[width=40pc]{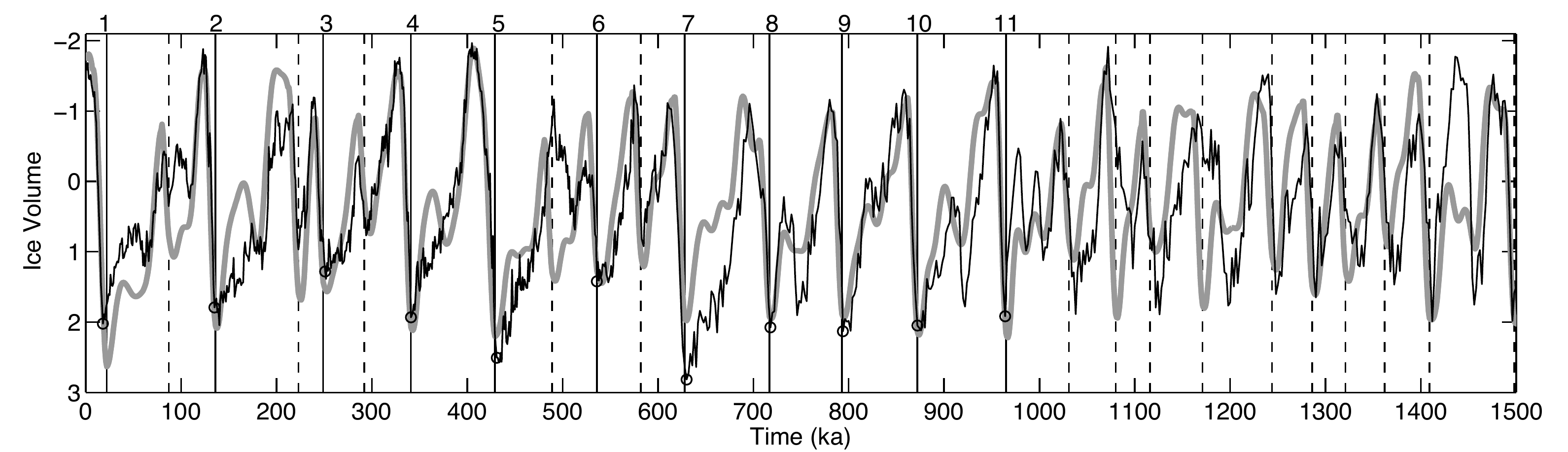}
\end{adjustwidth}
\caption{Phase-space model (thick gray) and detrended benthic $\delta^{18}$O
  stack (thin black). Terminations 1-11 are labeled; termination
  onsets in $\delta^{18}$O are marked with circles, and the corresponding model
  onsets are marked with solid vertical lines. Other model
  terminations are marked with dotted lines.}
\label{model1500}
\end{figure}

\begin{figure} 
\noindent
\center
\includegraphics[width=28pc]{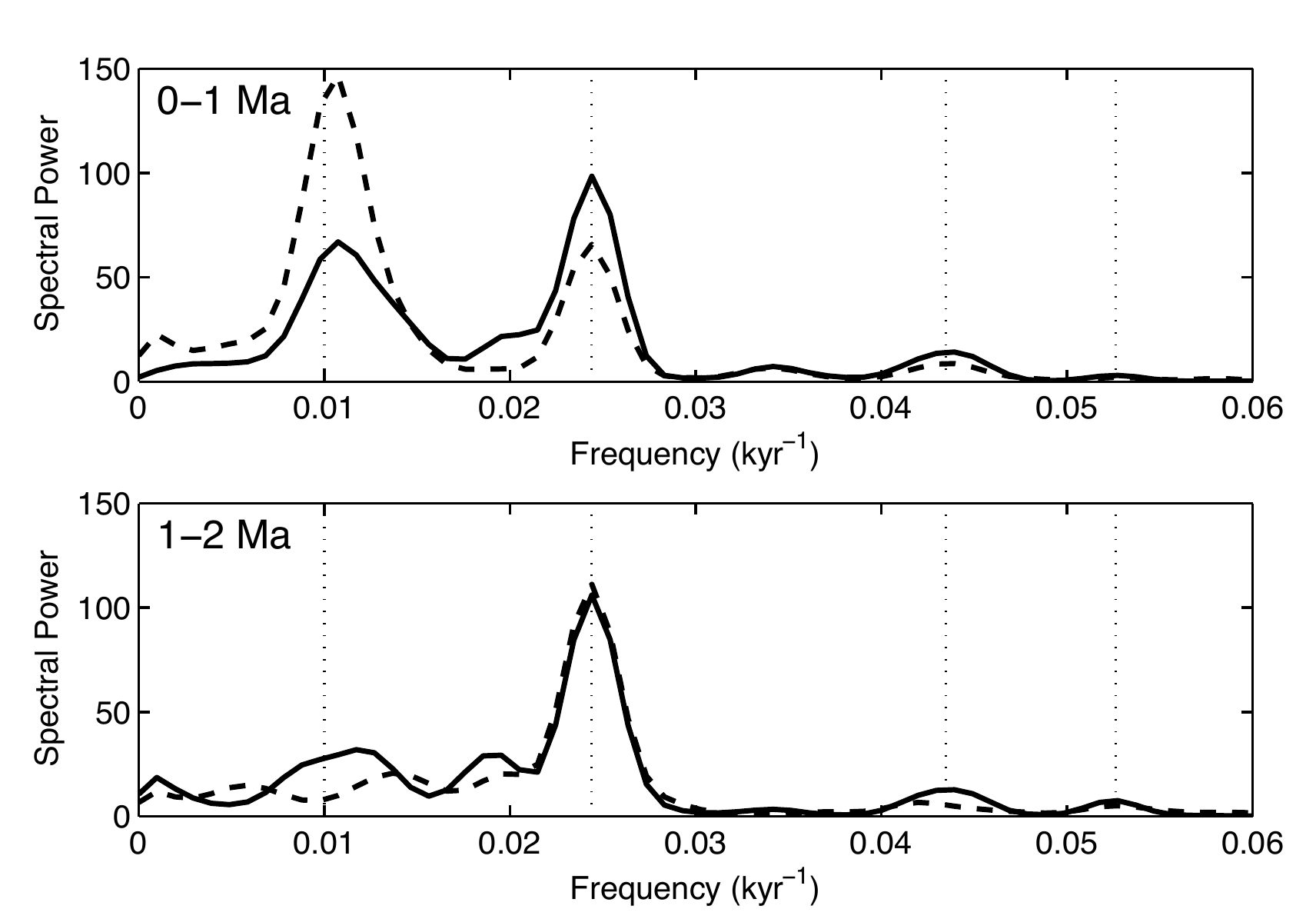}
\caption{Spectral power of model (solid) and detrended benthic $\delta^{18}$O
  stack (dotted) for 0--1 Ma (upper panel) and 1--2 Ma (lower
  panel). Dotted lines reference orbital frequencies corresponding to 100, 41, 23, and 19 kyrs. Spectral analysis uses Welch's method with 400-kyr windows
  and 75\% overlap. Model and data are scaled to have a standard
  deviation of 1 from 0--2 Ma.}
\label{spectra}
\end{figure}

The model reproduces many features of the detrended stack from 1.5--0
Ma (Figure \ref{model1500}), especially the ages and amplitudes of glacial
terminations of the last 1 Myr. The model also reproduces a shift from
41-kyr cycles to 100-kyr cycles at approximately 1 Ma. However, the
change in 100-kyr power is not as dramatic in the model as the proxy
data. In the model, power in the 100-kyr band (78.8--128.0 kyr)
increases from 17\% of the response before 1 Ma to 29\% after, whereas
100-kyr power in the data increase from 8\% to 48\% of the response
(Figure \ref{spectra}).

The most notable model-data misfits in the last 1.5 Myr occur during
MIS 5c at $\sim$100 ka, MIS 13 at $\sim$500 ka (which has also been
problematic in other models \citep{Pai98,Par03}), MIS 35 at $\sim$1175 ka, and MIS 47 at $\sim$1440
ka. (Interestingly, a much better match for MIS 47 is obtained if the
model is started with a value of 0 at 1.5 Ma.) The model also
generates five extra terminations in the last 600 ka (dotted vertical lines in Figure \ref{model1500}). 
However, three of the five extra
terminations have small amplitudes and well match the observed climate
response. The other two during MIS 5 and 13 actually self-correct the
model after it over estimates mid-interglacial ice growth. In fact,
recent sea level estimates suggest that ice volume may have been
smaller during MIS 5a than MIS 5c \citep{Dor10}; therefore,
some of the apparent misfit between model and during MIS 5 may be
caused by the temperature component of the $\delta^{18}$O stack.

\subsection{Phases of 100-kyr terminations}
\label{phases}
\begin{figure} 
\noindent
\begin{adjustwidth}{-2cm}{-2cm}
\center
\includegraphics[width=36pc]{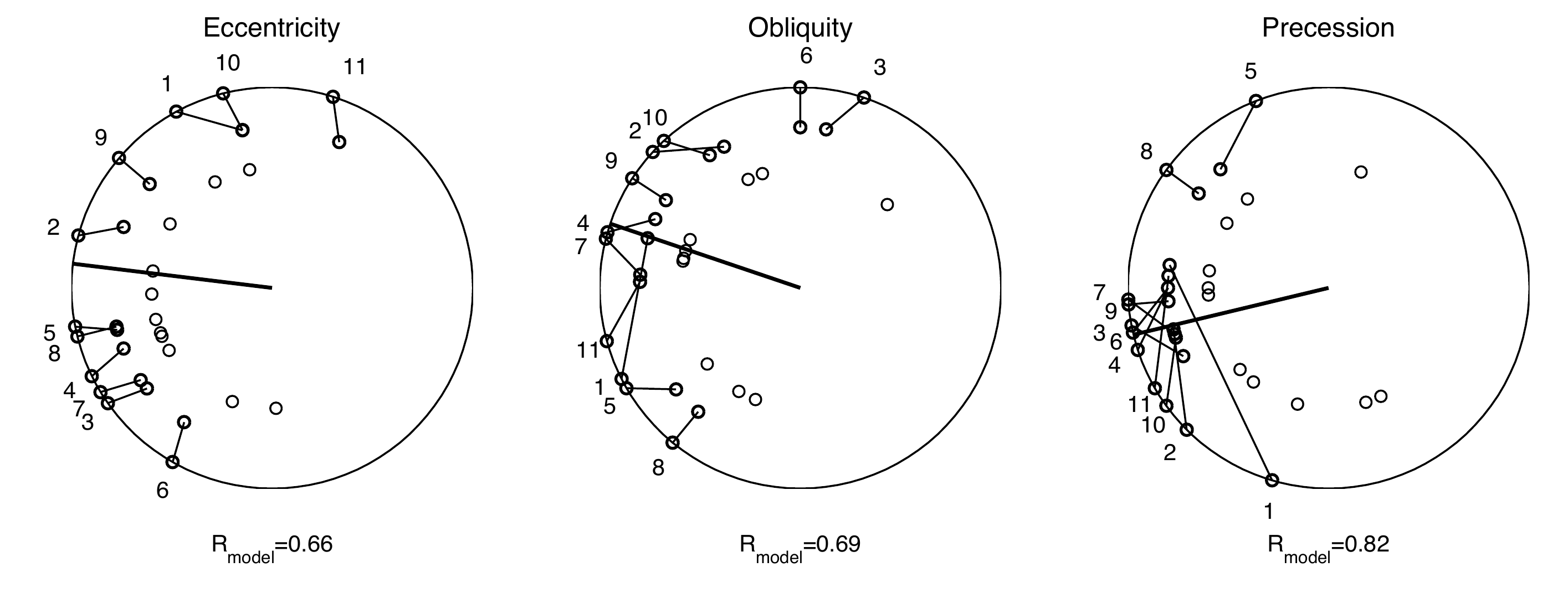}
\end{adjustwidth}
\caption{Phase wheels for terminations 1--11. The phase of model
  terminations are plotted and labeled on the outermost portion of the
  phase wheels. Tops of the wheels represent maxima in eccentricity
  (left) and obliquity (center) and minimum precession (corresponding
  to maximum northern hemisphere summer insolation, right panel). Lags
  increase clockwise. Model termination phases are connected to the
  corresponding terminations in the tuned $\delta^{18}$O stack (middle set of
  circles). The innermost set of circles shows termination phases of
  the untuned $\delta^{18}$O stack \citep{Lis10b}.}
\label{phase_wheel}
\end{figure}

Next we analyze the instantaneous orbital phases of Terminations 1-11
in the model and data. (Note that this does not include the five extra
model terminations discussed above.) The phase of termination onset is
analyzed rather than termination midpoint because the onset is
considered to reflect the threshold at which forcing causes a change
in the mode of climate response. The termination onset is explicitly
determined within the model.  Termination onsets in the data are
defined as the start of rapid $\delta^{18}$O increase (circles in Figure
\ref{model1500}). Figure \ref{phase_wheel} shows the phase of individual terminations. In both the
model and data, the average phase of termination onset is
approximately 90$^\circ$ before the maximum for all three orbital
cycles (where precession maximum is defined according to the maximum
in Northern hemisphere insolation). The phases of individual
terminations range from a lead of 164$^\circ$ to a lag of 25$^\circ$.

\subsection{Spectral Power}
\label{spectralpower}
Although cycle-for-cycle agreement between the model and data is
weaker before 1.5 Ma (Figure \ref{model3000}), the model reproduces important
spectral characteristics of the data for the last 3 Myr.  Both model
and data are dominated by 41-kyr power with very little 23-kyr and
100-kyr power from 2--1 Ma, and both have considerably more 100-kyr
power after 1 Ma (Figure \ref{spectra}). Although the model has a smaller increase
in 100-kyr power than the data, it achieves this without any change in
model parameters; the different model results for the two time
intervals are solely the result of differences in orbital
forcing. 

\begin{figure} 
\noindent
\center
\includegraphics[width=28pc]{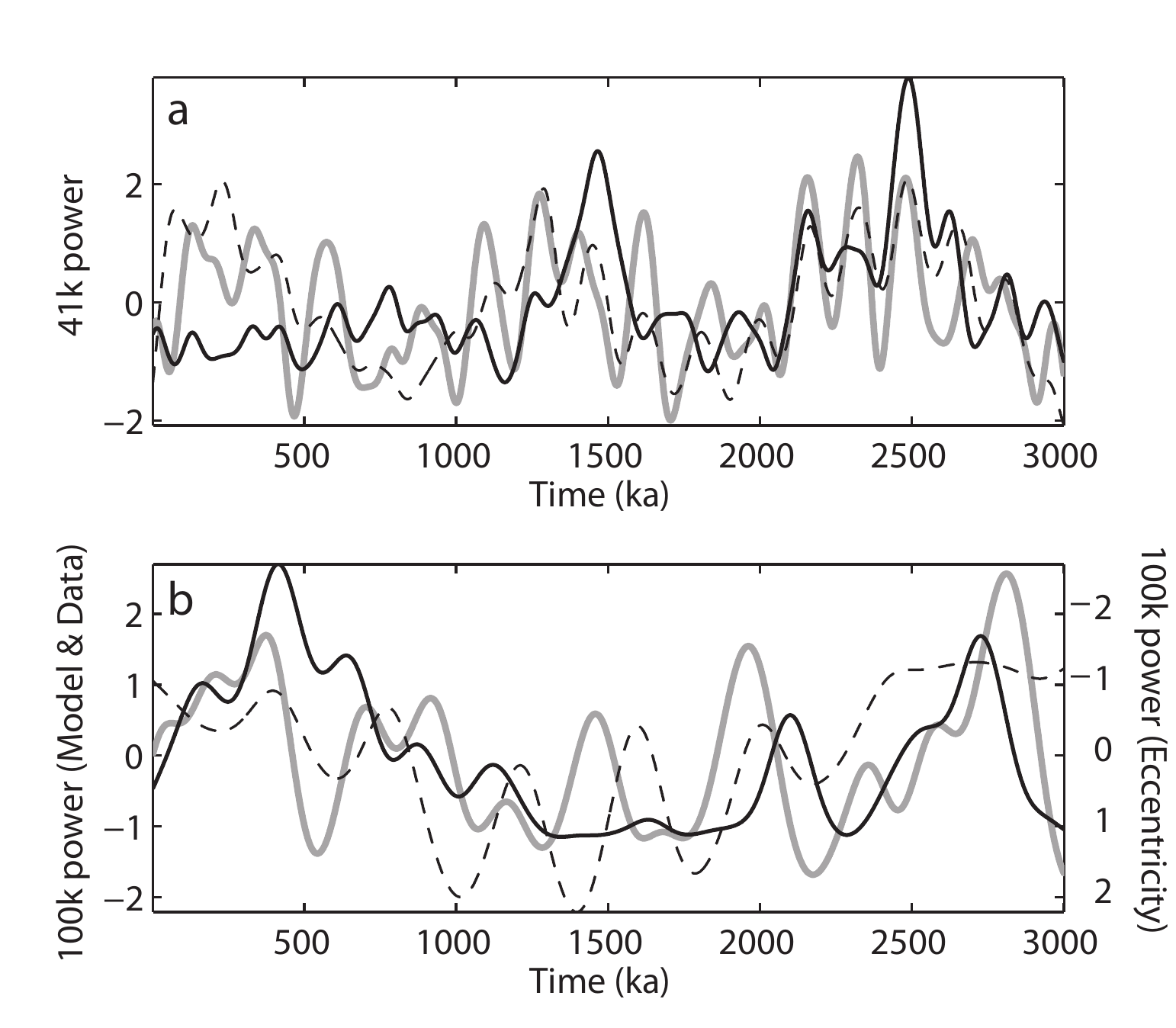}
\caption{Wavelet spectral power versus time in a) the 41-kyr band
  (36.8--46.4 kyr) and b) 100-kyr band (78.0--123.8 kyr). Spectral
  power for model (gray), data (black), and orbital parameters
  (dotted) are each scaled to have zero mean and unit standard
  deviation. The 100-kyr power of eccentricity (dotted) is flipped
  vertically to illustrate its negative correlation with model and
  data.}
\label{model_data_pwr_time}
\end{figure}

\begin{figure} 
\noindent
\begin{adjustwidth}{-2.5cm}{-2.5cm}
\center
\includegraphics[width=40pc]{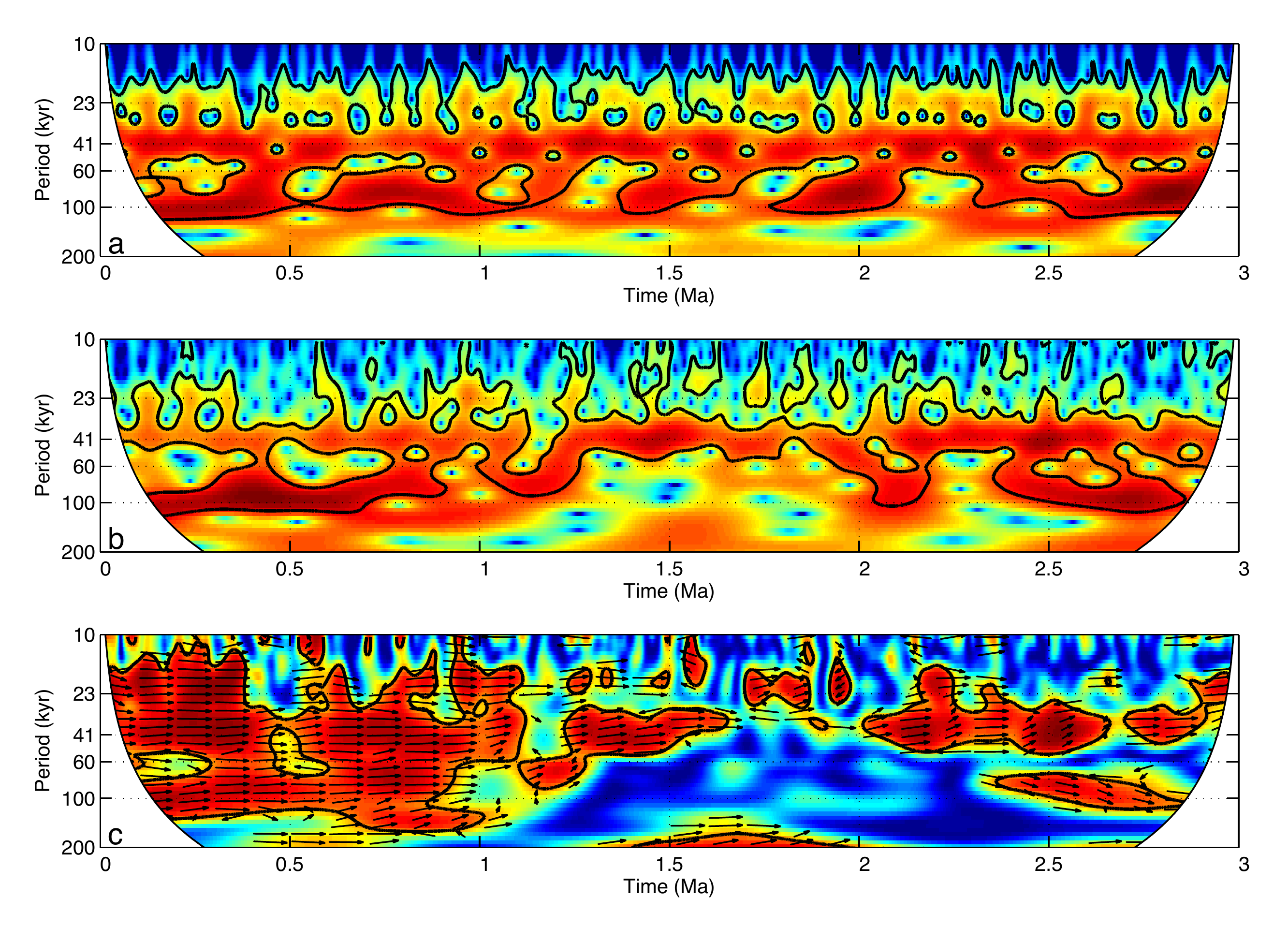}
\end{adjustwidth}
\caption{Wavelet spectral power of a) model and b) detrended benthic
  $\delta^{18}$O stack. c) Coherence between model and data. Arrows to the right
  show in-phase responses.}
\label{wavelet}
\end{figure}

Figure \ref{model_data_pwr_time} compares changes in 41-kyr and 100-kyr power of the model
and detrended stack over the last 3 Myr based on wavelet power spectra
\citep{Gri04}. Power in the 41-kyr band (35.3--46.5 kyr) of
the model and data correlate with the amplitude modulation of
obliquity before 0.7 Ma. After 0.7 Ma the 41-kyr power of the data no
longer correlates with the amplitude of obliquity forcing \citep{Lis07}. Power in the 100-kyr band (78.8--128.0 kyr) of
model and data are in good agreement for the entire 3 Myr and are both
negatively correlated with the 100-kyr power of eccentricity. This
feature of the data is described by \citet{Lis10b} and \citet{Mey10} and is discussed more in Section \ref{midpleistocenetransition}. Figure \ref{wavelet} shows
the complete wavelet power spectra of the model and detrended $\delta^{18}$O
stack and their coherence. Note the similaritiy of the transition from
41- to 100-kyr power in the model and data from 1.3--0.7 Ma. The
model's 100-kyr response is coherent with the data during the Late
Pleistocene and during a peak in 100-kyr response from 3.0--2.5
Ma. Both of these time intervals correspond to minima in the 100-kyr
power of eccentricity.

\section{Discussion}
\label{discussion}
\subsection{Model parameters}
\label{modelparameters}
Our model uses orbital forcing which is a linear combination of $\varepsilon$ (obliquity), $e\sin \omega$ (precession) and $e \cos \omega$ (phase-shifted precession). The choice of which linear combination to use is data-driven, in that a regression analysis is used to determine which combination best explains $y'$, the rate of change in ice volume. In our view, this is preferable to working with a preconceived notion of which latitude and season is the most suitable for use in the forcing function. In addition, by using weighted regressions, we allow the coefficients in the linear combination to depend on the ice volume $y$. 
Figure \ref{2a2b} shows that the weightings of the components of the forcing do vary significantly with ice volume.
We find that precession is more important during glacial epochs and obliquity is more important during interglacials---see Figure \ref{2a2b} (left). As the northern hemisphere ice margin moves southward, so does the latitude of the insolation curve whose obliquity and precession components match our modelÕs forcing function. 
This fits with the idea that the amount of melting in the most southerly parts of the ice sheets should be the primary driver of ice volume changes.
The phase of the precession forcing is also of interest. While the standard midsummer precession curve works well in interglacials, we find that as the ice volume grows, the relevant season for precession forcing moves 1-2 months towards spring---see Figure \ref{2a2b} (right).
This fits with the idea that as ice sheets extend further south, the solar intensity necessary for significant melting will be attained earlier in the spring.

We have also used a data-driven approach to determine the approximate shape of the internal evolution function (drift function) $g(y)$. This function specifies the forcing-independent part of the rate of change in ice volume in the accumulation state. It is significant that $g$ has an ``S'' shape (see Figure \ref{g}) so that horizontal lines intersect the curve in either 1, 2, or 3 points. Each horizontal line equates to a particular level of orbital forcing. Rising intercepts correspond to unstable fixed points, and descending intercepts correspond to stable fixed points. Thus for some values of the forcing, both glacial and interglacial states are stable. For higher values of the forcing, only the interglacial state is stable, and for lower values only the glacial is stable. 
Our model's drift function lies mostly below the axis, which means that on average, the ice volume grows (until it reaches the threshold for terminations). 
This scenario, with either one or two stable equilibria depending on insolation forcing, is firmly rooted in the physics of ice sheets \citep{Wee76}.
Other authors have used cubic nonlinearities like this in climate models, including
\citet{Sal93}; \citet{Dit09} has made use of ÒSÓ-shaped drift functions to reinterpret and extend Paillard's 3-state model \citep{Pai98} from the point of view of bifurcation theory.
It is worth comparing the shape of $g$ (assuming zero focing, as in Figure \ref{g}) with that of the corresponding function in \citet{Imb80}, which used straight lines with corresponding to time constants of 10.6 kyr (warming) and 42.5 kyr (cooling). If one replaces the portion of the graph in Figure \ref{g} below the axis with a straight line to the point (2,-.07), one obtains equivalent time constants of 12 kyr (warming) and 58.5 kyr (cooling).

While individual terminations can be sensitive to small changes in parameters, the overall pattern of 100-kyr cycles at times of low eccentricity is stable to small changes in parameters. We would expect something similar if the model were subjected to stochastic forcing---this would be interesting to pursue further. However, it is worth noting that a deterministic model can explain many of the key features of the climate record.

Larger changes in parameters can make the system either more or less likely to produce 100-kyr cycles. But it is important to note that our model is realistic in the sense that parameters were chosen close to those determined from the statistical analysis---see Figures \ref{2a2b} and \ref{g}.

\subsection{Termination phases}
\label{terminationphases}
The phase of termination onset with respect to orbital forcing has
been shown to vary when the forcing is not perfectly periodic \citep{Tzi06} and when the trigger for termination onset
depends on both ice volume and forcing \citep{Par03}. The results of our model suggest that the range of
termination phases produced by the climate system could be
approximately 180$^\circ$, twice as large as range described by \citet{Tzi06}. Therefore, ``early'' terminations may be
consistent with Milankovitch forcing as long as they occur after the
minimum in insolation forcing. These findings support the
interpretation of \citet{Kaw07} that the onset of abrupt
Antarctic warming is consistent with northern hemisphere forcing
during each of the last four terminations despite their wide range in
phase.

Phase stability with respect to different orbital parameters can be
quantified by Rayleigh's R statistic, defined as

\begin{equation}
  R=\frac{1}{N} \bigl\vert \sum_{i=0}^n \cos \phi_n + i \sin \phi_n \bigr\vert
\label{R}
\end{equation}

\noindent where $\phi_n$ is the phase lag of the $n$th 100-kyr window
relative to eccentricity and the line brackets indicate the
magnitude. R has a maximum value of 1.0 when all cycles have the same
phase lag. For Terminations 1--11 in the model and data, termination
phases have a slightly tighter clustering with respect to precession
than obliquity or eccentricity (Figure \ref{phase_wheel}). The precession R values of
the model and tuned data are 0.82 and 0.93 respectively, whereas the
obliquity R values are 0.69 and 0.72. It is unclear whether this is a
real feature of climate dynamics or an artifact of age model tuning
because the phase at which terminations occur in the tuned data will
affect the climate's apparent sensitivity to obliquity and precession
forcing and, therefore, the parameterization of the model. The
precession phase of terminations in the untuned stack \citep{Lis10b}
is slightly less clustered than for eccentricity and obliquity, but
this could easily be an age model artifact caused by the greater
impact of age model uncertainty on the phase of a shorter cycle
\citep{Huy05}.

\subsection{Mid-Pleistocene Transition}
\label{midpleistocenetransition}
Based on the anticorrelation between 100-kyr power in eccentricity and
the LR04 $\delta^{18}$O stack, \citet{Lis10b} proposed that the internal
climate feedbacks responsible for Plio-Pleistocene 100-kyr cycles are
inhibited by strong precession forcing. Our model demonstrates that
relatively simple responses to orbital forcing can reproduce the
observed anticorrelation between the 100-kyr power of eccentricity and
ice volume. The 100-kyr cycle in the presence of weak eccentricity (and
thus weak precession) occurs because the model is biased toward ice
volume growth between terminations and because the sensitivity to
precession increases as ice volume increases. Thus, after $\sim$90 kyr
of ice sheet growth, ice volume is large enough that the combined
influence of obliquity and relatively weak precession will trigger a
termination. In contrast, during strong 100-kyr eccentricity cycles
the combined obliquity and precession forcing is strong enough to
trigger terminations approximately every 41 kyr, as observed 1.5--1 Ma
and 600--450 ka. Additionally, positive and negative interference
between precession and obliquity might contribute to changes in the
100-kyr power of the model through time.

Before 1 Ma the timing of individual $\sim$100-kyr cycles in the model
does not correspond particularly well with $\sim$100-kyr cycles in the
data. The model also produces slightly too much 100-kyr power at 2.0
and 1.5 Ma and not enough at 500 ka. It is possible that a different
parameterization or a more sophisticated model might
alleviate these discrepancies.
However, the model is not capable of producing the trend and the reduction in amplitude that one sees in the data prior to 1 Ma. A more complete understanding of climate dynamics prior to 1.5 Ma is likely to require a model incorporating gradual changes in climate dynamics. 
Nevertheless, our model demonstrates a mechanism by which
100-kyr power in eccentricity could be anticorrelated with the climate
response. Additionally, it supports the hypothesis that the timing of
the MPT is related to changes in eccentricity forcing and suggests
that if any changes in climate dynamics or boundary conditions are
associated with the MPT, they may be relatively subtle and/or
gradual. For example, no significant trend in CO$_2$ concentrations
across the MPT is suggested by most paleoclimate records \cite{Lis10a}.

\section{Conclusions}
\label{conclusions}
A pair of evolution equations describe how climate responds to orbital forcing. A threshold in phase space determines which equation applies at any given time. When a combination of ice volume and its melting rate is not too large, the system is in an accumulation state that evolves according to a first-order differential equation. The system responds linearly to orbital forcing and is subject to a nonlinear drift term having an ``S'' shape.  The drift term leads to either one or two stable values for ice volume, depending on the level of orbital forcing. Above the threshold, the system is in a runaway melting or termination state that evolves according to a second-order equation. In phase space, the system moves along a semicircular arc toward a forcing-dependent destination with low ice volume, at which point it returns to the accumulation state. 

Our model uses a flexible forcing function that allows the strength of obliquity and precession forcing and the precession phase to vary with ice volume. The form of the forcing function is guided by statistical analysis of correlations between each orbital function and the rate of change of ice volume.

The model reproduces many of the features of the ice-volume record that have been challenging to understand.  The model has a ``100-kyr'' mode where climate stays in the accumulation state through one or more insolation peaks before reaching the termination threshold. This mode tends to be disrupted by high eccentricity values, which lead to larger insolation peaks and premature crossing of the termination threshold. Conversely, an extended period of low eccentricity (such as the cycle prior to stage 11) allows for an extended time in the accumulation state before the threshold is crossed, and this will lead to a larger 100-kyr cycle. Thus the model provides an explanation for the observed anticorrelation between eccentricity and the 100-kyr component of climate \citep{Lis10b}.  

Model output matches up well with many features of the LR04 $\delta^{18}$O stack from 1.5--0 Ma, especially the ages and amplitudes of glacial terminations of the last 1 Myr. Broad features of the power in the 100-kyr band match up with the data from 3--0 Ma---in particular the pulses of 100-kyr power at 400-kyr eccentricity minima and the periods of more intense 100-kyr power coinciding with broad low-eccentricity epochs at .5 and 2.5 Ma (the 2-Myr eccentricity cycle). Model output switches from 40-kyr mode to 100-kyr mode at about 1 Ma, thereby reproducing
the increase in 100-kyr power during
the mid-Pleistocene transition without any change in model parameters. This supports the hypothesis that Pleistocene variations in the 100-kyr power of glacial cycles could be caused,
at least in part,
 by changes in Earth's orbital parameters, such as amplitude modulation of the 100-kyr eccentricity cycle.

\bibliographystyle{model2-names}

%%\bibliography{phase.bib}
\newpage
%% not want to use model2-names.bst.



%% \bibitem must have one of the following forms:
%%   \bibitem[Jones et al.(1990)]{key}...
%%   \bibitem[Jones et al.(1990)Jones, Baker, and Williams]{key}...
%%   \bibitem[Jones et al., 1990]{key}...
%%   \bibitem[\protect\citeauthoryear{Jones, Baker, and Williams}{Jones
%%       et al.}{1990}]{key}...
%%   \bibitem[\protect\citeauthoryear{Jones et al.}{1990}]{key}...
%%   \bibitem[\protect\astroncite{Jones et al.}{1990}]{key}...
%%   \bibitem[\protect\citename{Jones et al., }1990]{key}...
%%   \harvarditem[Jones et al.]{Jones, Baker, and Williams}{1990}{key}...
%%


% \end{thebibliography}

\end{document}